\documentstyle[emulateapj,epsf,apjfonts]{article}

\slugcomment{Accepted for publication in the Astrophysical Journal}

\begin{document}

\title{Far-UV Emissions of the Sun in Time: Probing Solar Magnetic Activity and
Effects on Evolution of Paleo-Planetary Atmospheres\footnote{Based on
observations made with the NASA-CNES-CSA Far Ultraviolet Spectroscopic
Explorer. FUSE is operated for NASA by the Johns Hopkins University under NASA
contract NAS5-32985.}}

\author{Edward F. Guinan\altaffilmark{2}, Ignasi Ribas\altaffilmark{3},
and Graham M. Harper\altaffilmark{4}}

\altaffiltext{2}{Department of Astronomy \& Astrophysics, Villanova University,
Villanova, PA 19085, USA; E-mail: edward.guinan@villanova.edu}

\altaffiltext{3}{Departament d'Astronomia i Meteorologia, Universitat de
Barcelona, Av. Diagonal, 647, E-08028 Barcelona, Spain; E-mail: iribas@am.ub.es}

\altaffiltext{4}{Center for Astrophysics and Space Astronomy, University of
Colorado, 389 UCB, Boulder, CO 80309, USA; E-mail: gmh@casa.colorado.edu}

\begin{abstract}
We present and analyze {\em Far Ultraviolet Spectroscopic Explorer} (FUSE)
observations of six solar analogs. These are single, main-sequence G0--5 stars
selected as proxies for the Sun at several stages of its main-sequence lifetime
from $\sim$130 Myr up to $\sim$9 Gyr. The emission features in the FUSE
920--1180 \AA\ wavelength range allow for a critical probe of the hot plasma
over three decades in temperature: $\sim10^4$~K for the H~{\sc i} Lyman series
to $\sim6\cdot10^6$ K for the coronal Fe~{\sc xviii} $\lambda$975 line. Using
the flux ratio C~{\sc iii} $\lambda$1176/$\lambda$977 as diagnostics, we
investigate the dependence of the electron pressure of the transition region as
a function of the rotation period, age and magnetic activity. The results from
these solar proxies indicate that the electron pressure of the stellar
$\sim$10$^5$-K plasma decreases by a factor of $\sim$70 between the young,
fast-rotating ($P_{\rm rot}=2.7$ d) magnetically active star and the old,
slow-rotating ($P_{\rm rot}\sim35$ d) inactive star. Also, we study the
variations in the total surface flux for specific emission features that trace
the hot gas in the stellar chromosphere (C~{\sc ii}), transition region (C~{\sc
iii}, O~{\sc vi}), and corona (Fe~{\sc xviii}). The observations indicate that
the average surface fluxes of the analyzed emission features strongly decrease
with increasing stellar age and longer rotation period. The emission flux
evolution with age or rotation period is well fitted by power laws, which
become steeper from cooler chromospheric ($\sim10^4$~K) to hotter coronal
($\sim10^7$ K) plasma. The relationship for the integrated (920--1180 \AA) FUSE
flux indicates that the solar far-ultraviolet (FUV) emissions were about twice
the present value 2.5 Gyr ago and about 4 times the present value 3.5 Gyr ago.
Note also that the FUSE/FUV flux of the Zero-Age Main Sequence Sun could have
been higher by as much as 50 times. Our analysis suggests that the strong FUV
emissions of the young Sun may have played a crucial role in the developing
planetary system, in particular through the photoionization, photochemical
evolution and possible erosion of the planetary atmospheres. Some examples of
the effects of the early Sun's enhanced FUV irradiance on the atmospheres of
Earth and Mars are also discussed.
\end{abstract}


\section{Introduction}

The Sun's magnetic activity is expected to have greatly decreased with time
(Skumanich 1972; Simon, Boesgaard, \& Herbig 1985; Dorren \& Guinan 1994;
Guinan, Ribas, \& Harper 2002) as the solar rotation slows down because of
angular momentum loss in the stellar wind and resultant reduction in magnetic
dynamo-related activity. The study of the young Sun's far-ultraviolet (FUV)
fluxes using solar proxies provides important diagnostics for the state of the
younger Solar System and the physics of the much more active early Sun. The
comprehensive ``Sun in Time'' project, begun in 1988 (Dorren \& Guinan 1994),
focuses on the study of the long-term evolution of the outer atmosphere of an
early G star, from the zero-age main sequence to the terminal-age main
sequence. A crucial component of the program is a carefully-selected sample of
nearby solar analogs (G\"udel, Guinan, \& Skinner 1997) with spectral types
confined between G0-5~V and with well-determined physical properties (including
temperatures, luminosities, and metal abundances). We have obtained extensive
photometry ($UBVRI$) of these stars from ground-based observatories for over a
decade to determine their rotation periods, investigate starspots and possible
activity cycles. In addition, we have been able to estimate the stellar ages by
making use of their memberships in clusters and moving groups, rotation
period--age relationships, and, for the older stars, fits to stellar evolution
models. 

The sample of solar proxies within the ``Sun in Time'' program (see Guinan,
Ribas \& Harper 2002) contains stars that have masses close to 1~M$_{\sun}$ and
cover most of the Sun's main sequence lifetime from $\sim$130 Myr to $\sim$9
Gyr. Basically, the program stars have similar convective-zone depths and
chemical abundances to those of the Sun and vary only by their age and rotation
periods ($P_{\rm rot}$), and hence dynamo-generated magnetic activity, i.e.,
chromospheric, transition region (TR), and coronal emissions. In essence, we
use these one solar mass stars with $P_{\rm rot}$ between 2.7 and $\sim$35 days
as laboratories to study and test solar and stellar dynamo theories by varying
essentially only one parameter: rotation period.

The ``Sun in Time'' is a comprehensive and multi-frequency program that addresses
a variety of topics: study of short- and long-term magnetic evolution; physics
and energy transfer mechanisms of the chromosphere, transition region (TR) and
corona; evolution of the spectral irradiances of the Sun and their effects on
paleo-planetary environments and atmospheres. To this end we utilize
observational data spanning almost the entire electromagnetic spectrum, which
allows us to probe the structure of stellar/solar atmospheres. The FUV has
not been readily accessible until the successful launch of the {\em Far
Ultraviolet Spectroscopic Explorer} (FUSE). FUSE allows a critical probe of the
hot plasmas over nearly three decades in temperature: e.g., $\sim10^4$~K for
the H~{\sc i} Lyman series (Ly$\beta$, Ly$\gamma$, ...) through O~{\sc vi}
$\lambda\lambda$1032, 1038 at $\sim3\cdot10^5$~K, and the recently identified
Fe~{\sc xviii} $\lambda$975 coronal emission feature at $\sim6\cdot10^6$~K.
Several of the strongest emission features, such as C~{\sc iii}
$\lambda\lambda$977,1176 and O~{\sc vi} $\lambda\lambda$1032,1038, originate in
TR plasmas, and are pivotal for understanding the mechanisms of chromospheric
and coronal heating. Thus, spectrophotometry with FUSE fills the energy gap
between IUE/HST (1160--3200~\AA) and EUVE (80--720~\AA) to yield a complete and
comprehensive picture of a solar-type star's atmosphere at different ages and
rotation periods.

In this paper we present and analyze FUSE observations of six bright solar
analogs that span a wide range of rotation periods (and ages) from 2.7 days
($\sim$130 Myr) to $\sim$35 days ($\sim$9 Gyr). The stars, whose properties are
listed in Table \ref{tabprop} along with the Sun's, are: EK Dra, $\pi^1$~UMa,
$\kappa^1$ Cet, $\beta$ Com, $\beta$ Hyi, and 16 Cyg A. The present study
focuses on two particular aspects of the ``Sun in Time'' program that can be
effectively addressed with FUV observations: {\em 1.)} The flux ratios between
the C~{\sc iii} $\lambda$1176 and C~{\sc iii} $\lambda$977 lines (both within
the FUSE spectral range) yield empirical measures and estimates of the electron
pressure ($P_e$) of the TR; {\em 2.)} Integrated fluxes for selected emission
lines can be used to study the evolution of the solar FUV irradiances at
different ages and rotation periods.

\placetable{tabprop} 

\section{Observations}\label{secobs}

FUV observations of the six targets discussed here were obtained with FUSE. The
FUSE instrument consists of four spectroscopic channels (LiF1, LiF2, SiC1, and
SiC2) that cover a combined wavelength range 905--1187~\AA.  Detailed
information on the FUSE mission and its on-orbit performance are provided by
Moos et al. (2000) and Sahnow et al. (2000). 

The observations of the targets were secured during FUSE Cycles 1, 2 and 3.
Spectra of three stars ($\kappa^1$~Cet, $\beta$~Com, and $\beta$~Hyi) were
taken as part of the Guest Investigator program A083 (P.I. Guinan), one star
($\pi^1$~UMa) was observed as part of the program B078 (P.I. Guinan), and to
additional stars (EK Dra and 16 Cyg A) were acquired within program C102 (P.I.
Guinan). The large aperture (LWRS) was employed in all cases to minimize photon
losses due to telescope misalignments, and this caused a fairly severe
contamination of day-time spectra from geocoronal emission lines and scattered
solar radiation. Information on the dates and exposure times of the
observations is provided in Table \ref{taball}.

\placetable{taball} 

The FUSE data were retrieved from the Multimission Archive at Space Telescope
({\em MAST}), and then re-calibrated with CALFUSE
2.0.5\footnote{http://fuse.pha.jhu.edu/analysis/calfuse.html}.  For each star,
the time-tagged data for the sub-exposures were combined into a single dataset.
The spectral extraction windows were individually aligned to ensure correct
extraction of the stellar spectra. We employed the CALFUSE 2.0.5 event burst
screening and default background subtraction. The data were then calibrated and
screened separately for orbital night and day, allowing us to estimate any
contamination of the stellar spectra from ``airglow'' and scattered solar
light. The scattered light was found to be negligible compared to the stellar
source, except for the faint FUV target 16 Cyg A, which could suffer from some
contamination especially in the O~{\sc vi} features. 

The calibrated spectra are oversampled so they were re-sampled by binning over
four wavelength intervals, thereby increasing the signal-to-noise ratio. A
comparison of the emission line fluxes from different channels, e.g., C~{\sc
iii} 977~\AA\ SiC2A and SiC1B, provides an independent check on data artifacts,
and whether the star was in both apertures for the same length of time.

\section{Analysis} \label{secan}

\begin{figure}[!ht]
\epsscale{0.90}
\plotone{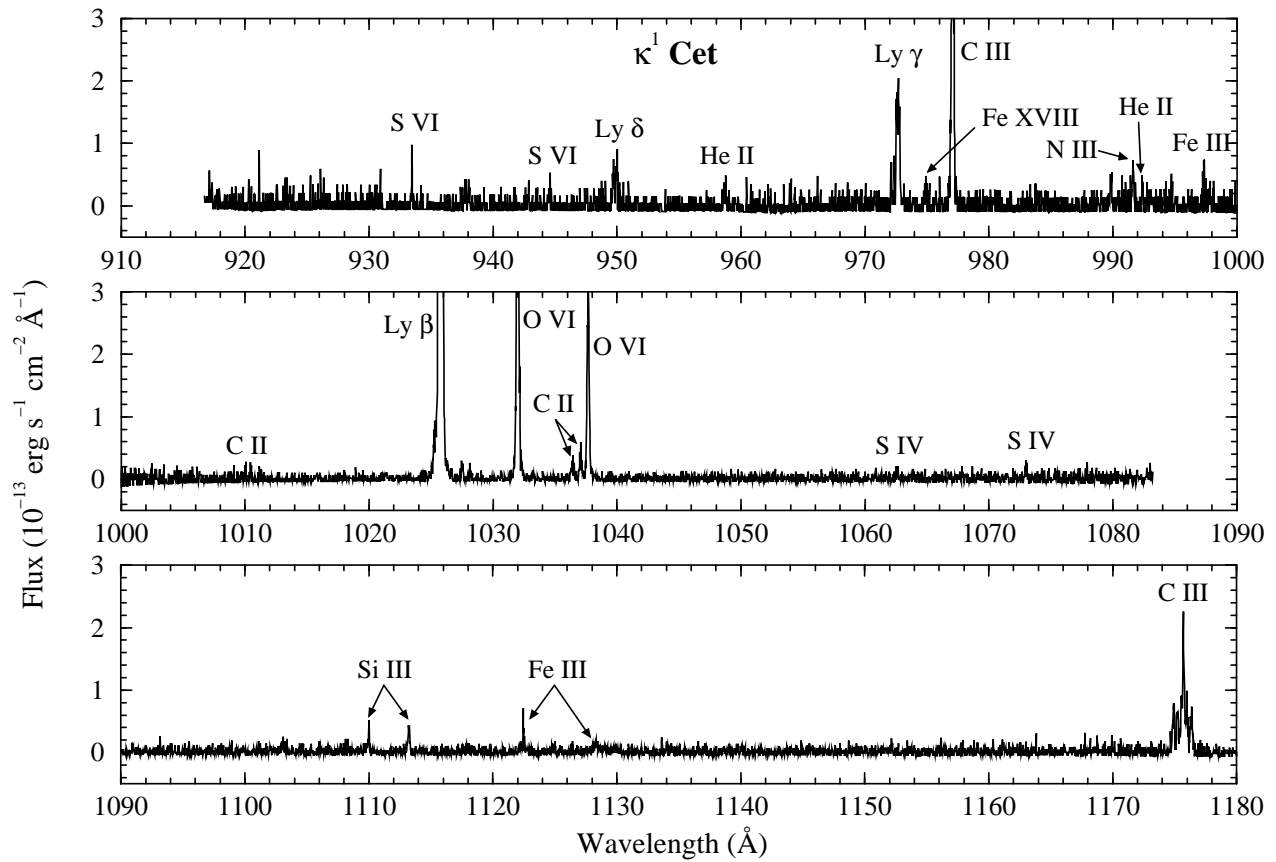}
\figcaption[f1.eps]{SiC2A (top), LiF1A (middle) and LiF2A (bottom)
night-time spectra of the target star $\kappa^1$ Cet (see Table \ref{tabprop}
for information on the stellar properties). The most prominent emission
features are labeled with the name of the corresponding ion. \label{figsp}}
\end{figure}

For illustration we show the extracted spectrum of $\kappa^1$ Cet in Figure
\ref{figsp}. Note that we only show the night-time data that presents less
contamination by terrestrial airglow and geocoronal emission. Thus, all
features in Figure \ref{figsp} are of stellar origin, although the H~{\sc i}
Lyman series still has some geocoronal contamination. The FUV spectra of the
other four targets are qualitatively similar, and only the spectrum of EK Dra
and $\pi^1$ UMa are somewhat noisier. For these two stars the increased noise
arises from the failure of the SiC2A and LiF2A observations and the consequent
use of the less sensitive segment B of detector 1 (SiC1B and LiF1B) for those
parts of the spectra. As can be seen in Figure \ref{figsp}, the continuum level
is negligible and the flux is in the form of emission lines.  Individual
channels are represented in the figure because the FUSE effective area is
heavily dominated by only one channel at each wavelength. In this case, SiC2A,
LiF1A, and LiF2A spectra are shown in the upper, central, and lower panels,
respectively. 

\placefigure{figsp}

Studies of FUSE observations of late-type stars, AB Dor and Capella, have been
presented by Ake et al. (2000) and Young et al. (2001), respectively. The
authors carried out thorough investigations of the chromospheric and TR
emission features in the FUSE wavelength range and provided identification
spectra. A comparison with our sample spectrum in Figure \ref{figsp} indicates
a striking resemblance to these spectra, except for small to moderate changes
in the relative strengths of some features. We have thus been able to identify
the most significant features in our spectra by using the results of Ake et al.
(2000) and Young et al. (2001).  Among the strongest FUV emission lines in
coronal spectra of late-type stars are the C~{\sc iii} singlet at 977.020~\AA,
the C~{\sc iii} multiplet around 1176~\AA, and the O~{\sc vi} doublet at
1031.925~\AA\ and 1037.614~\AA. We focus here primarily on the analysis of
stellar features that are not significantly affected by geocoronal
contamination and day-time spectra can be used without any loss of accuracy. We
thus combined the night-time and day-time data by using a exposure-time
weighting scheme.

All of our FUSE spectra show clean profiles for these lines that permit
reliable integrated flux measurements, which were carried out by means of
NOAO/IRAF tasks. Different methods (plain addition, Gaussian fitting) were
employed to measure the integrated line fluxes, study the associated
uncertainties, to check the line profiles, and also to assess the possible
presence of interstellar absorption (see below for further discussion). The
total integrated fluxes for each star and their intrinsic uncertainties (random
errors) are provided in Table \ref{taball}.

The ratio of the fluxes for the C~{\sc iii} $\lambda$1176 and $\lambda$977
emission features is provided in Table \ref{tabR}. This ratio, which yields an
excellent electron pressure/density diagnostic, is the topic of discussion for
\S\ref{secne}. The ratio of the two O {\sc vi} features
($\lambda$1038/$\lambda$1032) is not provided in Table \ref{tabR} but it is in
all cases very close to 1:2. This is the value expected for an effectively-thin
plasma where the flux ratio is proportional to the ratio of the collision
strengths for the two transitions, which for O~{\sc vi} is very close to 1:2
(Zhang, Sampson, \& Fontes 1990).

\begin{figure}[!ht]
\epsscale{0.50}
\plotone{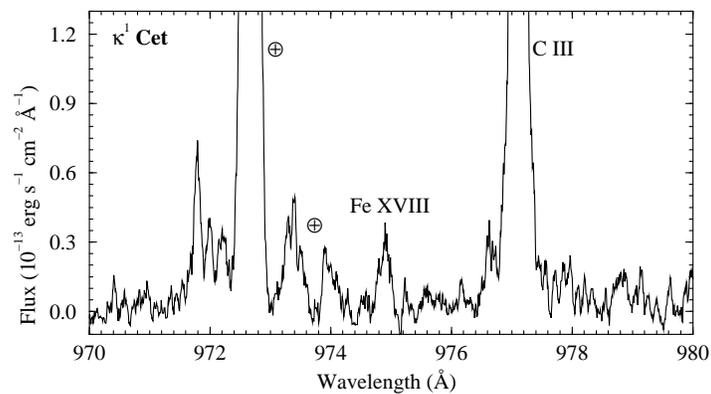}
\figcaption[f2.eps]{Detail of the FUSE spectrum of $\kappa^1$ Cet in
the wavelength region around the coronal Fe~{\sc xviii} emission feature. Also
present are the stellar C~{\sc iii} emission line and some ``airglow''
features. \label{figfe}}
\end{figure}

Plasma model fits to extensive X-ray observations of the three youngest stars
in the sample (EK Dra, $\pi^1$ UMa, and $\kappa^1$ Cet) indicate the presence
of high-temperature coronal material (G\"udel et al. 1997). As discussed by
Feldman \& Doschek (1991), Young et al. (2001) and Redfield et al. (2003),
coronal emission lines of highly-ionized ions arising from forbidden
transitions are expected in the FUSE wavelength range. A coronal Fe~{\sc xviii}
forbidden line at 974.85~\AA\ was identified for the first time in FUSE spectra
of Capella (Young et al. 2001).  FUSE spectra of EK~Dra, $\pi^1$ UMa and
$\kappa^1$~Cet also clearly show this emission feature (see Figure \ref{figfe}
for an example). Very recently, Redfield et al. (2002) have reported the
detection of a coronal Fe~{\sc xix} $\lambda$1118 emission line in the FUSE
spectra of several nearby cool stars.  The FUSE spectra of EK~Dra, $\pi^1$ UMa
and $\kappa^1$~Cet show weak Fe~{\sc xix} $\lambda$1118 features barely above
the noise level. As pointed out by Redfield et al. (2003), this line can be
corrupted by a nearby C~{\sc i} multiplet.  Because of the poor S/N of the flux
measurements for Fe~{\sc xix} $\lambda$1118 and concerns with nearby line
contamination have focused only on the clean Fe~{\sc xviii} $\lambda$975 line
as a tracer of the hot coronal plasma. Integrated flux measurements have been
carried out using the same method described above and are reported in Table
\ref{taball}. No trace of this feature is detected for the older stars in the
sample down to the noise level of the spectra, and flux upper limits are
therefore provided in Table \ref{taball}.

\placefigure{figfe}

In addition to integrating the fluxes, we also obtained line centroid positions
and FWHM measurements for the selected features in Table \ref{taball}. The
motivation for such measurements was the search for possible line shifts or
broadenings that could arise from plasma outflows or extended material. 
A critical point when measuring line shifts is an accurate wavelength scale.
Aperture centering problems caused by misalignments may result in systematic
velocity shifts among the different channels. Redfield et al. (2002) used ISM
features to derive the absolute wavelength scale of their spectra. In our case
ISM absorption features are not measurable and we followed a different
approach. Namely, we aligned the O~{\sc vi} $\lambda\lambda$1032,1038 doublet,
which is available in all four channels, using the LiF1 channel as the
reference (this is the channel used for guiding). This procedure should ensure
relative velocities accurate to 5--10 km~s$^{-1}$ in all channels. The velocity
corrections were generally quite relevant with values up to 40 km~s$^{-1}$.
With a corrected wavelength scale, we measured the line centroids and velocity
shifts for several features. The results are given in Table \ref{tabRV}. No
significant line shifts with respect to the photospheric velocity were observed
down to the precision of the measurements. The expected $\sim$10~km~s$^{-1}$
redshift of TR features (e.g., Redfield et al. 2002) is only observed in the
most active star of the sample, EK Dra. The rest of the stars actually exhibit
slightly blueshifted TR features, albeit with very low statistical
significance.

\placetable{tabRV}

FWHM measurements for all the emission lines in Table \ref{taball} yielded
relatively small widths of $0.20-0.25$~\AA, which are equivalent to velocities
of $60-70$~km~s$^{-1}$. Only EK Dra, the most active of our targets, was found
to have broader line profiles. In this case, the measured FWHM was
about 0.35~\AA\ or 110~km~s$^{-1}$. (The actual rotational velocity of this
star is $v_{\rm rot}\approx 17$~km~s$^{-1}$.) A closer inspection of the C~{\sc
iii} $\lambda$977 and O {\sc vi} $\lambda$1032 line profiles revealed extended
line wings in some of the targets. This is a well-known phenomenon (see, e.g.,
Linksy \& Wood 1994) that is usually modeled by assuming double Gaussians with
components of different widths. For the stars in our sample, only EK Dra shows
unambiguous evidence of a broad component, whereas $\pi^1$~UMa and $\kappa^1$
Cet show somewhat extended line wings but the two-component fits yielded
inconclusive results. Two-Gaussian fits to the O {\sc vi} $\lambda$1032 line
of EK Dra indicate a ratio between the broad component flux and the total flux
of 0.38. Thus, extended line profiles appear to correlate with stellar activity
(and age), in agreement with the results of Redfield et al. (2002) and Wood,
Linsky, \& Ayres (1997). The presence of broad and narrow components in
high-temperature lines is still not well understood. A plausible model (Wood et
al. 1997) suggests that these may arise from high-velocity nonthermal motions
during magnetic reconnection events (microflares).

\subsection{Interstellar absorption corrections} \label{secism}

Before analyzing and modeling the line integrated fluxes, the possible effects
of interstellar medium (ISM) absorption must be investigated. The C~{\sc iii}
$\lambda$977 line and the H~{\sc i} Lyman series lines are the most sensitive
to attenuation effects because of possible superimposed ISM absorption. As
shown in Table \ref{tabprop}, the stars in the sample are nearby and lie within
7--34 pc of the Sun and have negligibly small values of $E(B-V)$ from ISM dust.
However, even for the small ISM column densities at these close distances
($N_{\rm H}\sim10^{18}$ cm$^{-2}$; Redfield \& Linsky 2000), one expects some
absorption in the lines mentioned above because they are among the strongest
ISM features in the entire spectrum. Unfortunately, both the resolution and S/N
of our FUSE spectra are not sufficiently high for reliable direct estimations
of the (expected weak) ISM FUV absorptions. We therefore adopted an alternative
procedure and carried out indirect determinations of the flux corrections for
the individual spectral features.

To estimate the ISM column densities in the line of sight of our targets we
made use of the local ISM observations and model by Redfield \& Linsky (2002).
These authors analyzed high-resolution observations of Fe, Mg and Ca features
to determine column densities of these elements toward stars within 100 pc of
the Sun. One of our targets, $\kappa^1$ Cet, is included in Redfield \&
Linsky's study and thus direct ISM measurements are available. For the other
five targets we adopted the local ISM characteristics of neighboring stars with
direct measurements: DK~UMa for EK~Dra and $\pi^1$~UMa, HZ 43 for $\beta$~Com,
$\zeta$~Dor for $\beta$~Hyi, and $\delta$~Cyg for 16~Cyg~A. The C abundance was
computed from the observed column densities of Mg~{\sc ii} and Ca~{\sc ii} and
local ISM log abundances (relative to H) of $-3.66$, $-5.58$ and $-7.64$ for C,
Mg~{\sc ii}, and Ca~{\sc ii}, respectively, from Wood et al. (2002b) and
Redfield \& Linsky (2000). The C~{\sc iii} column density was subsequently
obtained by adopting the local ISM ionization ratio of C~{\sc iii}/C~{\sc
ii}$\approx$0.02 from Wood et al. (2002b). (Note that C~{\sc ii} is the
dominant ionization species in the ISM.)

The ISM absorption equivalent widths were estimated from the C~{\sc iii} column
densities and the transition oscillator strengths through the curve of growth
method (see Spitzer 1978). For the C~{\sc iii} $\lambda$977 transition, despite
its strength, the ISM absorption features for all our targets were found to be
relatively weak (unsaturated) because of the small column densities. Then,
using the ISM component velocities and the observed C~{\sc iii} line profile,
we calculated the fraction of stellar flux absorbed by the ISM. Our results
indicate that the flux corrections for the C~{\sc iii} $\lambda$977 line are
very small (3-7\%) for EK Dra, $\pi^1$ UMa, $\kappa^1$ Cet, and $\beta$ Com,
and somewhat larger (11-13\%) for $\beta$ Hyi and 16~Cyg~A. 

\subsection{The H~{\sc i} Lyman series and total FUSE fluxes} \label{lyman}

In addition to the features mentioned above, we attempted an estimation of the
total integrated fluxes within the FUSE bandpass (920--1180 \AA). However, even
when restricting to night-time spectra, contamination of the stellar H~{\sc i}
Lyman features by geocoronal emission is very significant. This is because FUSE
observations in this program were obtained through the large aperture to
ensure, where possible, that the important C~{\sc iii} 977~\AA\ emission line
was detected in the SiC channels. As a consequence, the Lyman geocoronal
emission is much broader than for the medium aperture, reaching about
$\pm100$~km~s$^{-1}$ near Ly$\beta$. This emission overlies and in most cases
completely dominates the stellar emission profile. The stellar signal is
expected to be a self-reversed emission feature with a dark ISM absorption
core.

To estimate the stellar fluxes for Ly$\beta$ we took the difference between the
night and day spectra, and assumed that the shape of geocoronal emission
remains constant. A fraction of this geocoronal emission was then subtracted
from the night spectra so that the residual flux at the location of the ISM
feature is consistent with the flux predicted by a simulation of the stellar
profile attenuated by the ISM. If emission peaks remained in the subtracted
spectrum we assummed that the remaining signal was stellar. Only $\kappa^1$~Cet
and $\pi^1$~UMa had any significant signal to satisfy these requirements. 
To estimate the shape of the underlying stellar emission profile, required to
estimate the total stellar flux, we used the functional form of the
``universal'' profile for strong partially coherent scattered lines derived by
Gayley (2002). This allowed us to use the observed flux in the wings, which is
relatively uncontaminated by geocoronal emission, to extrapolate towards line
center and provide a total flux estimate. The model stellar profile was then
attenuated by an ISM model (described above) and convolved with a PSF of $R\sim
15000-18000$ to mimic the FUSE spectral resolution. From the fits we obtained
total Ly$\beta$ fluxes of 4.0$\cdot$10$^{-13}$ erg~s$^{-1}$~cm$^{-2}$ and
2.4$\cdot$10$^{-13}$ erg~s$^{-1}$~cm$^{-2}$ for $\pi^1$~UMa and $\kappa^1$~Cet,
respectively. Given the uncertainty of the spectra, the adopted ISM model and
the assumed stellar profile shapes, we expect that the accuracy of estimated
stellar flux is no better than a factor of 1.5--2. Future observations of H
Ly$\alpha$ and MDRS observations of Ly$\beta$ are warranted and required to
improve the quantification of the total stellar FUV emission.

A correction to account for the rest of the H~{\sc i} Lyman series was derived
by comparison with SOHO SUMER spectra for the Sun. We roughly estimate that the
total flux contributed by the H~{\sc i} Lyman lines within the FUSE bandpass is
about 1.6 times the Ly$\beta$ flux. Then, the total flux in the range
920--1180~\AA\ was calculated by adding the contributions from the individual
features discussed above and the integrated H~{\sc i} Lyman series. We obtained
values of 10.5$\cdot$10$^{-13}$ erg~s$^{-1}$~cm$^{-2}$ and 8.7$\cdot$10$^{-13}$
erg~s$^{-1}$~cm$^{-2}$ for $\pi^1$~UMa and $\kappa^1$~Cet, respectively. Our
integrations indicate that the H~{\sc i} Lyman features in this wavelength
range (Ly$\beta$, Ly$\gamma$, etc) are important contributors ($\sim$40--60\%)
to the total flux of both $\pi^1$~UMa and $\kappa^1$~Cet. This is also
presumably the case of the remaining solar-type stars in the sample but the
weak stellar signal and strong geocoronal contamination of the FUSE
observations prevented us from carrying out reliable measurements.

\section{Emission line fluxes in the FUV}\label{secir}

One straightforward application of the measured line fluxes presented in Table
\ref{taball} is the analysis of their evolution along the main sequence of a
one solar mass star. To compare the absolute irradiances, we scaled
the observed fluxes to surface fluxes at a radius of 1~R$_{\odot}$. To estimate
the actual radius of each target we made use of the stellar luminosity
(computed from the Hipparcos distances and apparent magnitudes) and effective
temperature (obtained from spectroscopic measurements in the literature). The
radii of the targets are included in Table \ref{tabprop} and using these we
calculated the equivalent flux for a star of 1~R$_{\odot}$ in radius. As can be
seen in Table \ref{tabprop}, the radii of all stars but $\beta$ Hyi and 16 Cyg
A are very close to 1~R$_{\odot}$ (within 5-8\%) and the correction was
typically very small. The situation is somewhat different for $\beta$ Hyi
because of its advanced evolutionary stage and its mass being about
$\sim$1.1~M$_{\odot}$. In this case the flux correction was quite significant. 

\begin{figure}[!ht]
\epsscale{0.50}
\plotone{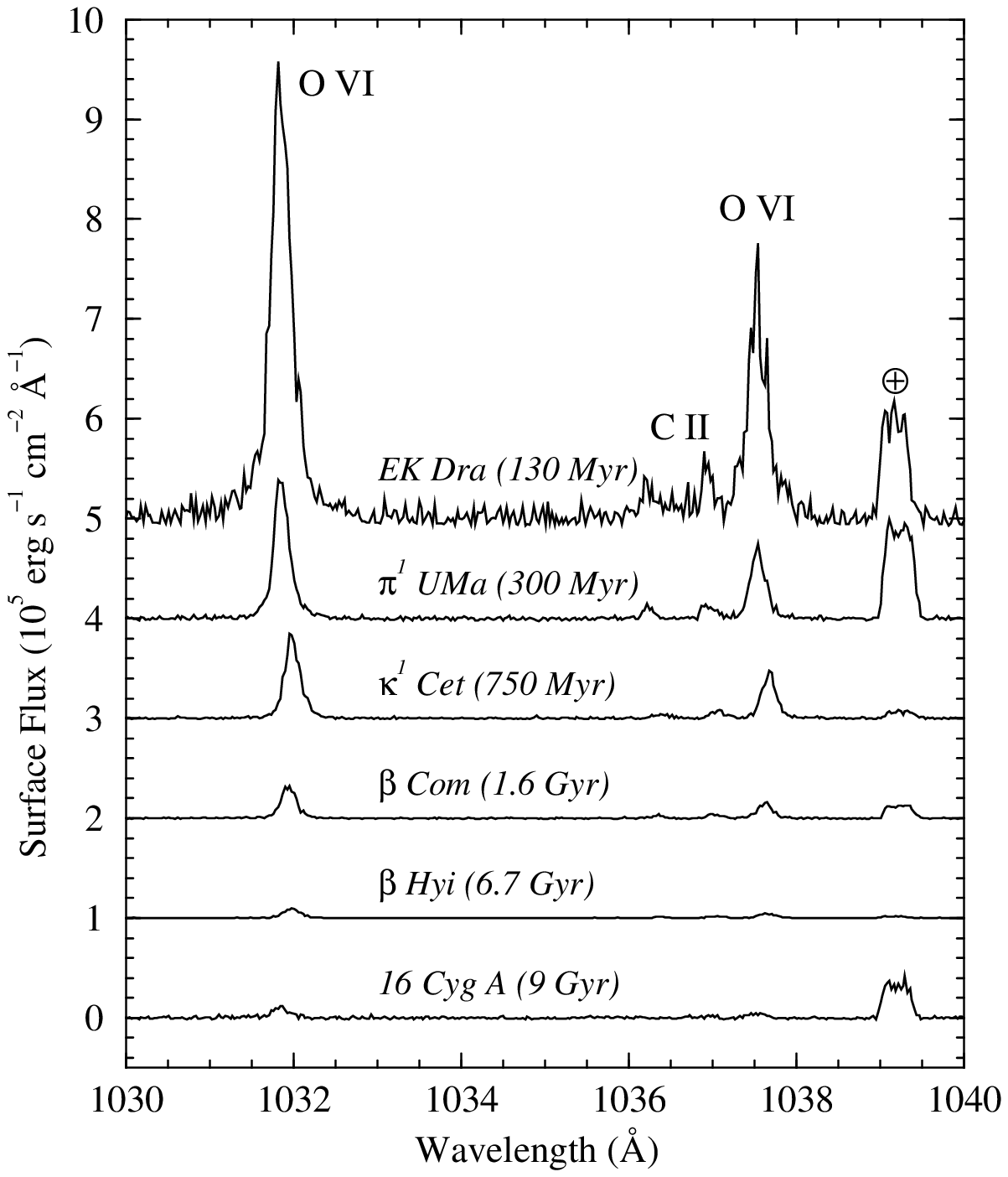}
\figcaption[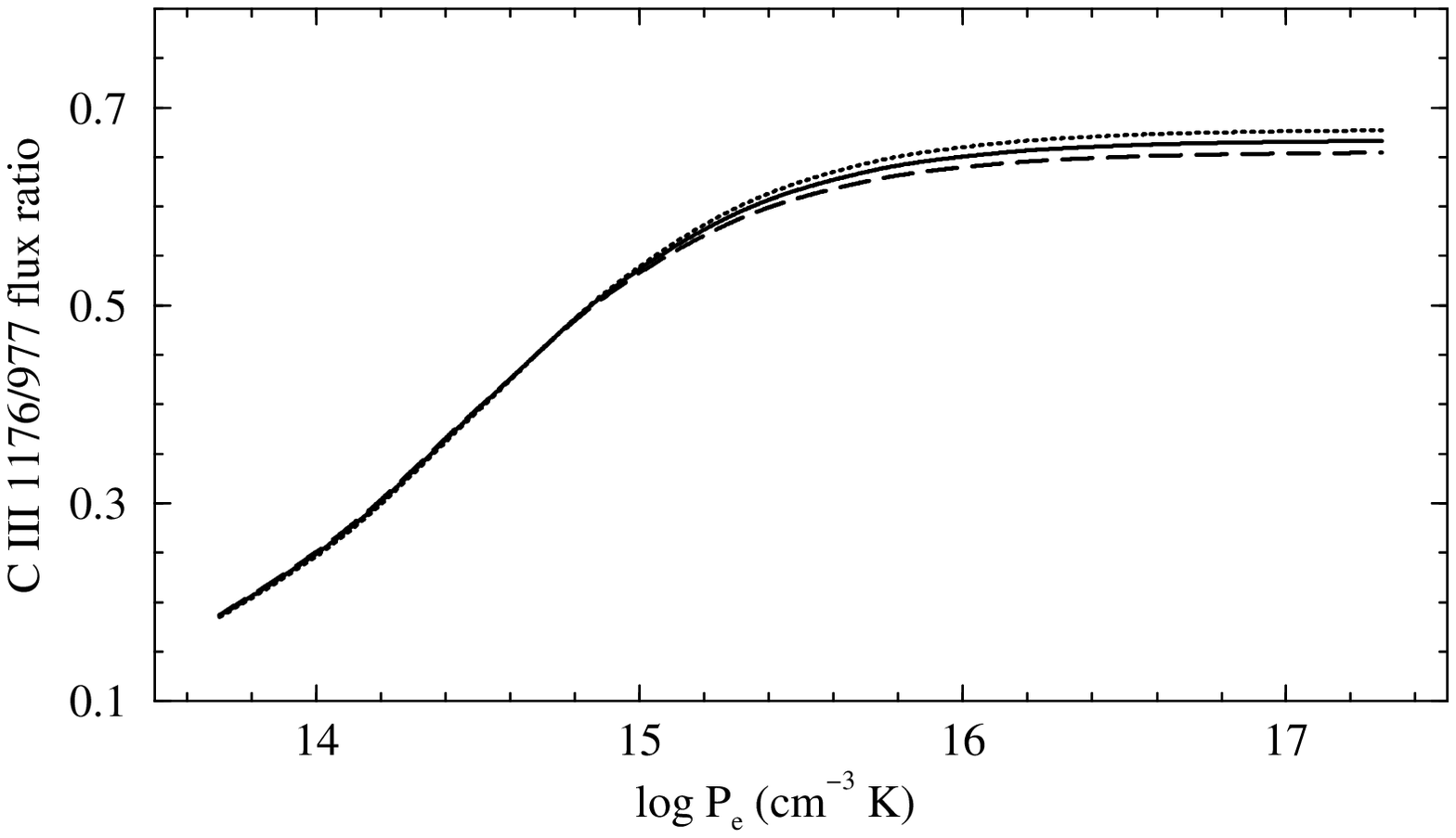]{Comparison of the surface fluxes for all targets in
our FUSE programs in the region around the O~{\sc vi} $\lambda\lambda$1032,1038
doublet. The spectra have been zeropoint-shifted using integer multiples of
$10^5$ erg~s$^{-1}$~cm$^{-2}$~\AA$^{-1}$ to avoid confusion. Note the obvious
trend of decreasing flux with increasing stellar age. The feature located near
$\lambda=1039$~\AA\ is of geocoronal origin. \label{figOvi}}
\end{figure}

The adopted final stellar surface fluxes for the four emission features studied
are provided in Table \ref{tabsurf}. Also in this table are the total
integrated FUV fluxes in the FUSE wavelength range (920--1180 \AA) for two of
the targets ($\pi^1$~UMa and $\kappa^1$~Cet). All fluxes have been corrected
for ISM absorption (see \S \ref{secism} and \ref{lyman}). Before proceeding
with a numerical analysis of these data, an illustrative visual impression of
the surface flux changes for the targets stars can be seen in Figure
\ref{figOvi}.  We have plotted details of the spectra (flux corrected to the
stellar surface) in a narrow 10-\AA\ wavelength interval that contains the
O~{\sc vi} doublet at $\lambda\lambda$1032,1038. A rough comparison of the line
variations as a function of stellar age does reveal a clear and definite trend
or decreasing strength with increasing age. 

\placetable{tabsurf}

\placefigure{figOvi}

\begin{figure}[!ht]
\epsscale{0.90}
\plotone{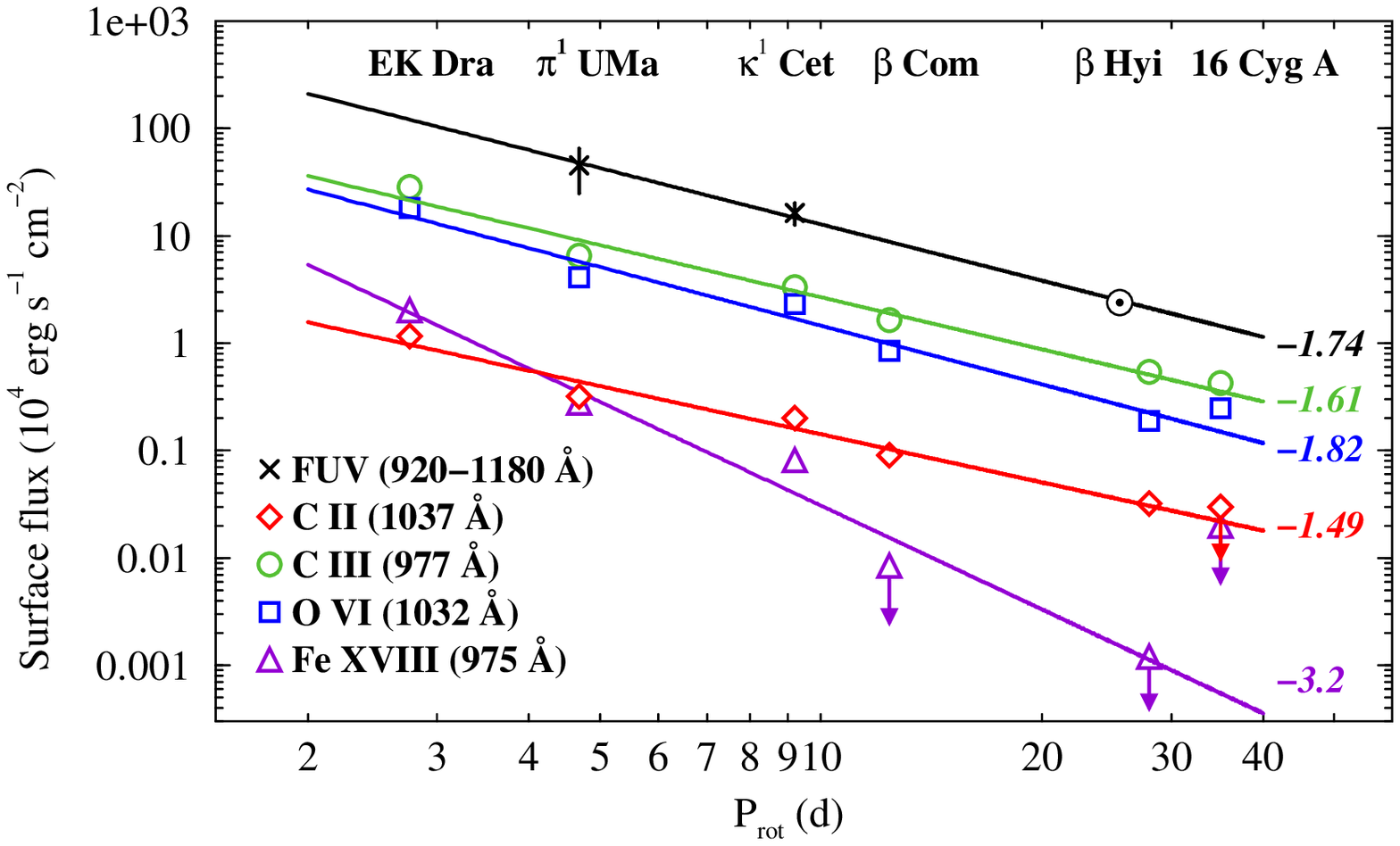}
\figcaption[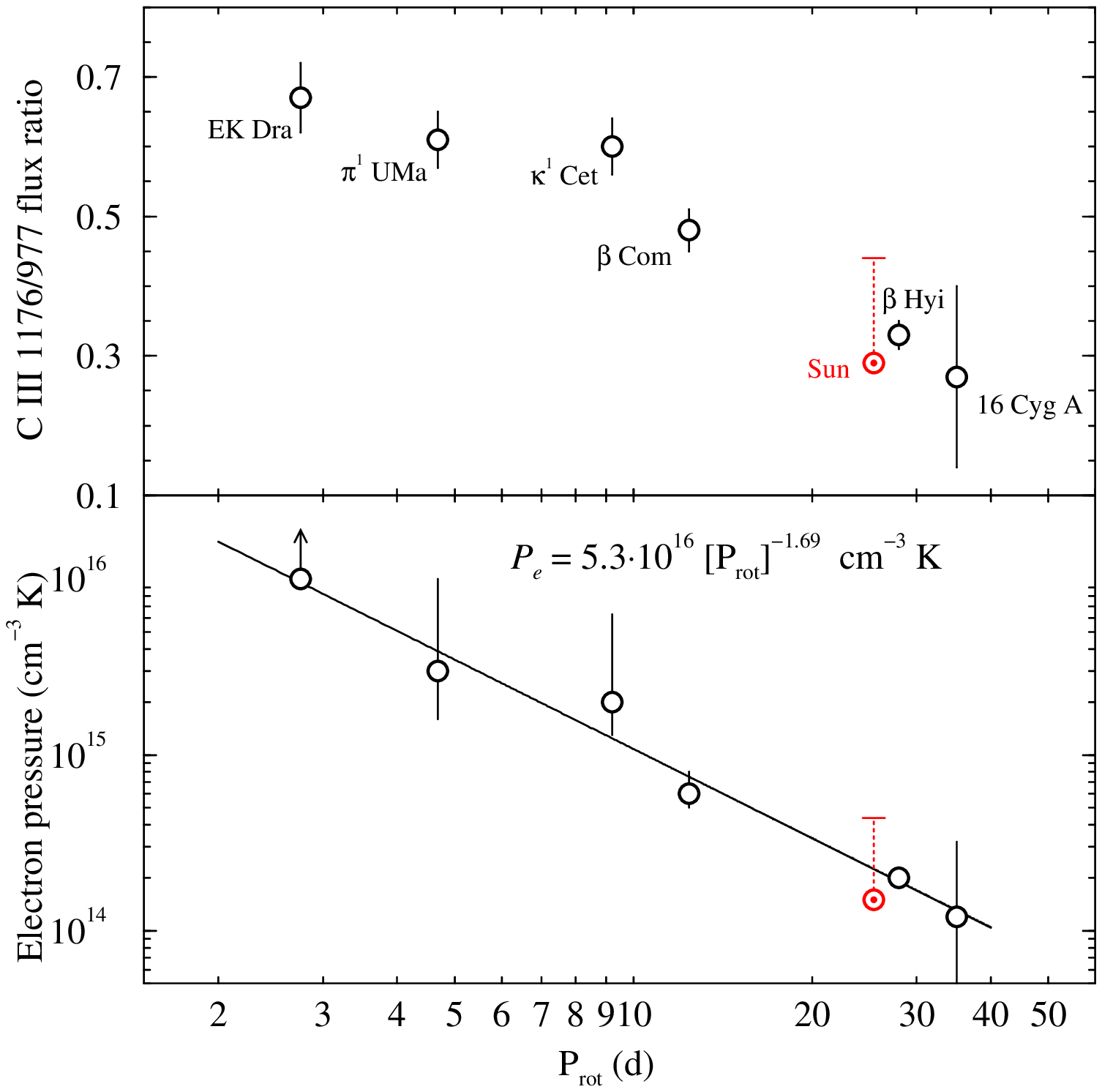]{ISM- and radius-corrected surface fluxes vs.  rotation
period for our six FUSE targets. Plotted here are flux measurements for four
features corresponding to transitions of C~{\sc ii}, C~{\sc iii}, O~{\sc vi}
and Fe~{\sc xviii}, which trace plasmas with temperatures of $20,000$~K,
$60,000$~K, $300,000$~K and 6 MK, respectively. Also included are measurements
of the integrated FUV (920--1180 \AA) flux in the FUSE wavelength range for two
of the targets. In this case, the integrated solar flux in the same wavelength
interval was used to constrain the fit (see text). The error bars of the
measurements are plotted but smaller than the size of the symbols in most
cases. The straight lines are power law fits to the measured data and the
corresponding slope coefficients are printed at the right end. \label{figIrr}}
\end{figure}

For a more quantitative analysis we used the integrated surface fluxes in Table
\ref{tabsurf}, which are plotted in Figure \ref{figIrr} as a function of the
stellar rotation period. The features represented are C {\sc ii} $\lambda$1037,
C {\sc iii} $\lambda$977, O {\sc vi} $\lambda$1032, and Fe~{\sc xviii}
$\lambda$975. The typical peak formation temperatures of the C~{\sc ii}, C~{\sc
iii}, O~{\sc vi}, and Fe~{\sc xviii} ions are approximately $20,000$~K,
$60,000$~K, $300,000$~K, and 6~MK, respectively (Arnaud \& Rothenflug 1985).
Thus, these lines probe a wide interval of plasma temperatures. As can be seen,
the newly-detected coronal line of Fe~{\sc xviii} is very important to our
investigation because it extends the coverage to the hot plasma component. Note
the much smaller surface fluxes of $\beta$ Hyi with respect to the younger
solar analogs for all the features studied. Roughly, $\beta$ Hyi's flux levels
are comparable to, or slightly smaller than, those of today's Sun. In addition,
Figure \ref{figIrr} also depicts the total integrated FUV fluxes in the FUSE
bandpass for $\pi^1$~UMa and $\kappa^1$~Cet.

\placefigure{figIrr}

The random measurement errors affecting the fluxes in Figure \ref{figIrr} are
discussed in \S \ref{secan}. However, additional systematic uncertainties can
be caused by the intrinsic stellar activity cycle. From an analogy with the
Sun, we expect peak-to-peak cycle amplitudes in the FUV of about 30--50\%.
Since the target stars are observed at different stages of their activity
cycles, these activity variations can be a source of scatter in the irradiance
plot. Fortunately, this activity-related scatter is rendered negligible by the
large relative differences between the fluxes of the targets (with factors of
over 30 in flux).

The fluxes of Figure \ref{figIrr} can be fit to good accuracy with power laws
of different slopes. These are listed in Table \ref{tabslopes}, which contains
the power-law slopes using both rotation period ($P_{\rm rot}$) and age as
independent variable. Note that only two FUSE measurements are available for
the total flux within 920--1180~\AA. To constrain the fit we used the flux
values for the quiet sun obtained through integration of the solar reference
spectrum by Heroux \& Hinteregger (1978) in the FUSE wavelength window. The
resulting surface flux was 2.4$\cdot$10$^4$~erg~s$^{-1}$~cm$^{-2}$.

Interestingly, the slopes not only change from line to line, but also they show
a clear trend. The power law becomes steeper as we move from cooler to hotter
material. The most extreme trend is that of the flux for the coronal ion
Fe~{\sc xviii}. The results indicate that the flux is reduced by a factor of
over 1000 when the rotation period increases ten-fold. To serve as a
reference, plasma of similar (or maybe slightly lower) temperature is probed
when analyzing fluxes in the X-ray domain. G\"udel et al. (1997) studied
X-ray fluxes (0.1--2.4 keV) of about a dozen solar analogs covering also a wide
period range. A power law fit to the data yielded a slope of $-2.64\pm0.12$,
which is in very good agreement with our value for the decrease of the Fe~{\sc
xviii} flux. 

\placetable{tabslopes}

The results indicate that the emission from hotter plasma decreases more
rapidly than that from cooler plasma. Analogous behavior was observed by
G\"udel et al. (1997) when comparing the emission measures derived from
observations in hard and soft X-rays. A possible scenario to explain the
steeper flux decrease for hotter plasma could be the weakening of the stellar
dynamo as the stars spin down with age and the corresponding decline in the
efficiency of heating mechanisms (e.g. flares) and in the strength of magnetic
fields that confine the gas.

It should be noted that our program stars, except for age and rotation period,
have similar properties to the Sun (mass, radius, effective temperature, metal
abundance). Importantly, these stars should all have similar convection zone
depths, which together with rotation, is an important parameter in most modern
magnetic dynamo theories. Thus, the stars in our sample can serve as
laboratories for testing the generation (and dissipation) of magnetic
dynamo-related energy and heating in solar-type stars where rotation (angular
velocity) is the only important variable. The observed power law dependencies
on rotation of the various emission line fluxes should be important as inputs
for testing and constraining stellar/solar dynamo theories.     

\section{Electron pressures in the transition region}\label{secne}

\begin{figure}[!ht]
\epsscale{0.50}
\plotone{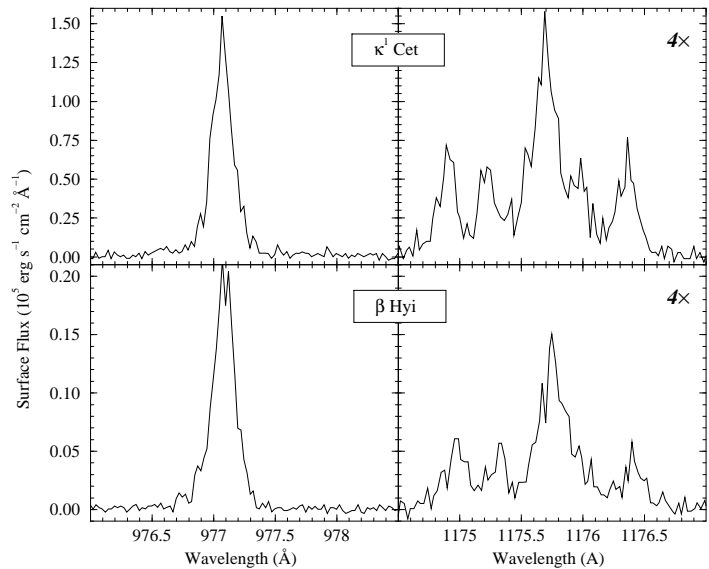}
\figcaption[f3.eps]{Details of the C~{\sc iii} $\lambda$977 and
$\lambda$1176 emission features in the $\kappa^1$ Cet and $\beta$ Hyi FUSE
spectra. The C~{\sc iii} $\lambda$1176 feature is a multiplet that comprises
six transitions. The flux scale in the C~{\sc iii} $\lambda$1176 panels has
been enhanced by a factor of 4 for illustration purposes. \label{figcomp}}
\end{figure}

The C~{\sc iii} emission line at 977~\AA\ and the C~{\sc iii} multiplet at
1176~\AA\ occur in the FUSE wavelength region. The ratio of the C~{\sc iii}
$\lambda$1176/$\lambda$977 emission line fluxes has long been recognized to be
a sensitive diagnostic for measuring the electron pressure ($P_e$) of hot
($\sim$60,000 K), optically-thin plasmas (e.g., Dupree, Foukal, \& Jordan 1976;
Keenan \& Berrington 1985). FUSE offers an excellent opportunity to exploit the
full potential of this important diagnostic (${\cal R}\equiv{\cal
F}^{\mbox{\tiny C{\sc iii}}}_{\mbox{\tiny $\lambda$1176}}/{\cal F}^{\mbox{\tiny
C{\sc iii}}}_{\mbox{\tiny $\lambda$977}}$) because both C~{\sc iii} features
can be measured simultaneously. Furthermore, unlike the UV C~{\sc iii}]
$\lambda$1909 line, these FUV features are not contaminated by the stellar
continuum or nearby line emissions. Details of the C~{\sc iii} $\lambda$977 and
$\lambda$1176 emission features for two of the targets in the sample are shown
in Figure \ref{figcomp}.

\placefigure{figcomp}

To compute ${\cal R}$ ratios we adopted a simple C~{\sc iii} model which
includes the first 10 fine-structure levels. The energy levels are taken from
Moore (1970), the highest level in the model being $2p^2\>^1S_0$ at
$18219.88\>{\rm cm}^{-1}$. The electric dipole oscillator strengths are taken
from the compilation of Allard et al. (1990), except for transitions calculated
by Tachiev \& Froese Fischer (1999; 2002, priv. comm.) and Fleming, Hibbert, \&
Stafford (1984). The electric quadrupole and magnetic dipole oscillator
strengths are from Nussbaumer \& Storey (1978) and Tachiev \& Froese Fischer
(1999) and the $2s2p\>^3P_2^o - 2p^2\>^1S_0$ and $2s^2\>^1S_0 - 2s2p\>^3P_2^o$
magnetic quadrupole strengths are from Shorer \& Lin (1977) and Tachiev \&
Froese Fischer (1999), respectively.  The electron collision strengths are
taken from Berrington (1985) and Berrington et al. (1985). The proton collision
rates for transitions within $2s2p\>^3P^o$ are from Ryans et al. (1998) and
within $2p^2\>^3P$ are from Doyle, Kingston, \& Reid (1980).  We solved the
statistical equilibrium equations, under the assumption that the plasma is in
steady state, with no photo-excitation from external sources. The ratios
presented here are computed under the assumption that all the C~{\sc iii}
transitions are optically thin. We have also computed the ratios under
optically thick conditions using escape probabilities with hydrogen columns of
$10^{20}\>{\rm cm}^{-2}$ and assuming solar abundances (Grevesse \& Sauval
1998). We find that ${\cal R}$ is quite insensitive to optical depth effects
under these conditions, in agreement with the findings of Bhatia \& Kastner
(1992).

The $\lambda$1176/$\lambda$977 flux ratio of Be-like C~{\sc iii} is well known
to be sensitive to both electron density and temperature and is therefore
sensitive to the {\em shape of the emission measure distribution} between $\log
T_e=4.5$ and 4.9. The density sensitivity arises from collisional
thermalization of the lower levels of the 1176 multiplet ($2s2p\>^3P^o$) which
connect to the ground state ($2s^2\>^1S$) by an intercombination electric
dipole and a weak magnetic quadrupole transition at 1908.7~\AA\ and 1906.7~\AA,
respectively. The contribution functions for 1176~\AA\ and 977~\AA\ contain the
product of the collisional excitation which increases with $T_e$, the
ionization balance of C~{\sc iii} which peaks near $\log T_e=4.85$ (Arnaud \&
Rothenflug 1985), and the shape of the emission measure distribution between
$\log T_e=4.5$ and 4.9 which is a strong declining function of $T_e$, e.g.,
Jordan et al. (1987). The peak contribution is expected to lie below $\log
T_e=4.85$. For example for the Sun, Judge et al. (1995) find a peak as low as
$\log T_e=4.5$ for the 977-\AA\ transition, while MacPherson \& Jordan (1999)
find $\log T_e=4.7$. For the solar analog transition regions, we consider that
the C~{\sc iii} emission lines are formed at a constant electron pressure, and
we derive the ratio ${\cal R}$ as a function of electron pressure rather than
at a (less realistic) single temperature.

The emission measure distribution at C~{\sc iii} temperatures may change as a
function of stellar magnetic activity. However, as pointed out by Jordan
(2000), both the analyses of main sequence stars by Jordan et al. (1987) and
the similarity of the C~{\sc iv}/C~{\sc ii} UV emission line ratios for a wide
range of active coronal stars (Oranje 1986), indicate that the shape of the
emission measure distribution in the C~{\sc iii} forming region approximately
follows the shape of the inverse of total radiative loss curve between $\log
T_e=4.3$ and 5.3. If the density sensitivity of the ionization balance and
abundance gradients can be neglected, then the relative electron pressures
derived for our sample should be reliable.

\begin{figure}[!ht]
\epsscale{0.50}
\plotone{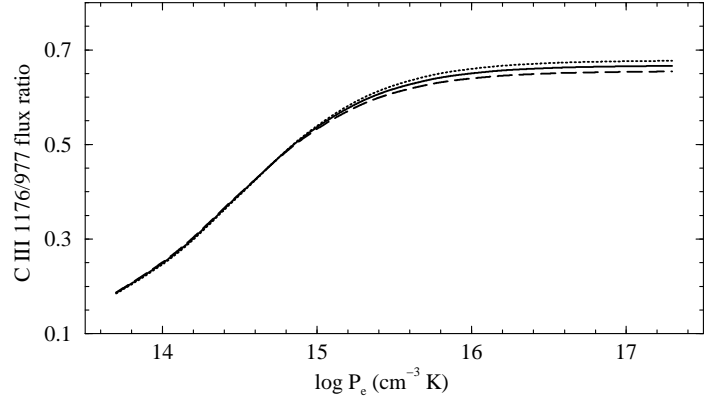}
\figcaption[f4.eps]{The flux ratio ${\cal R}$ for constant pressure models
with three differential emission measure distributions (DEM). In the C~{\sc
iii} formation region the DEM decreases with increasing $T_e$. The gradient of
the DEM increases from the top curve down; top (dotted line) $\alpha=1.5$,
middle (solid line) $\alpha=2.0$ and bottom (dashed line) $\alpha=2.5$ -- where
the DEM is given by Eq. \ref{eqDEM}.  The characteristic formation temperature
decreases with increasing $\alpha$ and gives rise to the to a lower ratio at a
given pressure. The difference in the curves is small compared to the
uncertainties in the observed ratio. \label{figPres}}
\end{figure}

To quantify the effects of assuming differential emission measure distributions
(DEM) with different slopes, we adopt a form
\begin{equation} \label{eqDEM}
{\rm DEM}\left(T_e\right)= n_e n_H {dr\over{d\ln T_e}} \propto T_e^{-\alpha}
\end{equation}
where $n_e$ and $n_H$ are the electron and hydrogen densities, respectively,
and $r$ is the radial coordinate. Here we assume that the shape of DEM is
similar to the emission measure over the C~{\sc iii} 977~\AA\ formation region.
Note that $\alpha \simeq 2$ is a reasonable approximation to the radiative
power-loss function for this temperature range.  For each pressure we integrate
over temperature from $\log T_e=4.3$ to 5.3, where the fractional abundance of
C~{\sc iii} exceeds 0.01, for each DEM to derive the total fluxes and derive
${\cal R}$. In Figure \ref{figPres} we show the ratio ${\cal R}$ for
$\alpha$=1.5, 2.0, and 2.5. This figure shows that the resulting ratio is not
too sensitive to uncertainties in the shape of the emission measure
distribution. Indeed, uncertainties in the collision strengths lead to greater
uncertainties in the electron pressure.

\placefigure{figPres}

\begin{figure}[!ht]
\epsscale{0.50}
\plotone{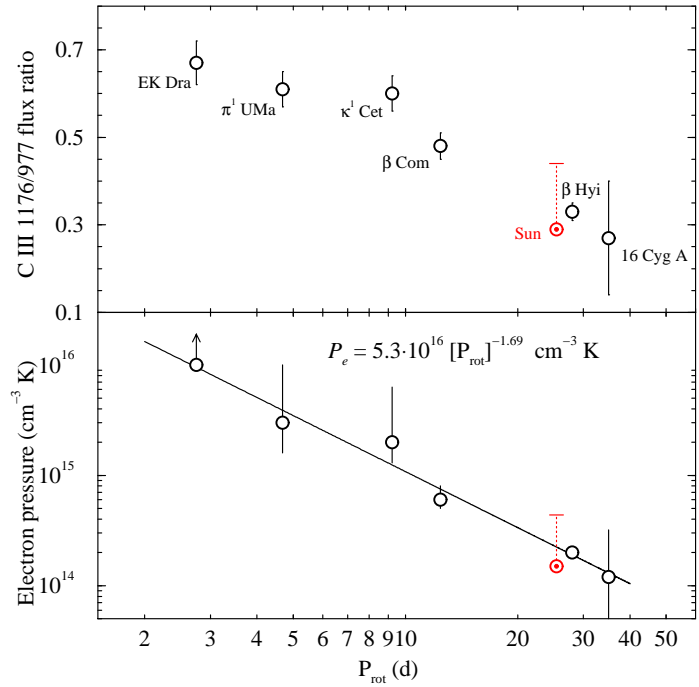}
\figcaption[f5.eps]{{\em Top:} Observed values of the flux ratio
${\cal R}\equiv$~C~{\sc iii} $\lambda$1176/$\lambda$977 as a function of the
stellar rotation period (which is an age indicator) for the stars in the
sample. The typical ${\cal R}$ values for the Sun (corresponding to quiet and
active regions) are also plotted. {\em Bottom:} Electron pressures derived from
the C~{\sc iii} $\lambda$1176/$\lambda$977 diagnostics using the theoretical
calculations in Figure \ref{figPres}. The straight line is a power law fit (all
data excluding the Sun) and the resulting expression is explicitly given in the
panel. \label{figRobs}}
\end{figure}

The empirical C~{\sc iii} ratios ${\cal R}$ computed for the six stars in the
sample are given in Table \ref{tabR}. Also provided are the ${\cal R}$ values
corrected for ISM absorption. From these one can estimate the electron
pressures through the relationship discussed above. When doing so, we obtain
$P_e$ values between $10^{14}$ and $10^{16}$ cm$^{-3}$ K, approximately. The
actual electron pressures computed for each of the stars in the sample are
given in Table \ref{tabR}. Both the ${\cal R}$ values and the derived plasma
electron pressures are plotted in in Figure \ref{figRobs} as a function of the
stellar rotation period. For comparison, the value for the quiet Sun was
determined to be ${\cal R}$=0.29 (Dupree, Foukal, \& Jordan 1976), and the
active Sun (measured in solar active regions such as sunspots) has a C~{\sc
iii} ratio of ${\cal R}$=0.44 (Noyes et al. 1985; Doyle et al. 1985). Note that
the ${\cal R}$ values determined for our stars correspond to the integrated
stellar disk average, thus including both active and inactive regions, weighted
towards region of large areal filling factor and high $n_e$. Interestingly, the
observed data for the target stars suggest a strong correlation between the
activity level of the star (= age or $P_{\rm rot}$) and ${\cal R}$ or $P_e$.

\placetable{tabR}

\placefigure{figRobs}

The elusive density of the TR has been probed for the first time along the
evolutionary path of a solar-type star, from ZAMS to TAMS. Our results show
that the electron pressure of the TR decreases monotonically as the star spins
down (and magnetic dynamo activity decreases). From the fit in Figure
\ref{figRobs} we infer a power-law relationship with a slope of $\sim-1.7$.
This suggests that the electron pressure of the solar TR may have decreased by
about a factor of $\sim$40 since the beginning of its main sequence evolution.
A likely explanation for the greater plasma electron pressure for the young and
active (rapidly rotating) stars is a stronger magnetic confinement of the
emitting material.

\begin{figure}[!ht]
\epsscale{0.50}
\plotone{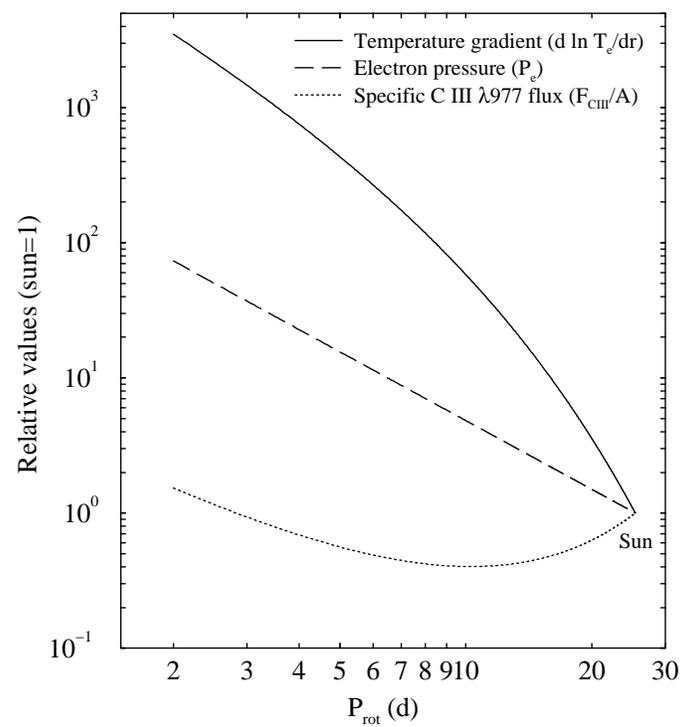}
\figcaption[f5.eps]{Variations in the temperature gradient, electron pressure
and specific C~{\sc iii} $\lambda$977 flux scaled to the solar values. The
derivation of the relationships plotted is discussed in \S \ref{secne}. 
Note the steep decrease in the temperature gradient with increasing rotation
period (or age) and the slowly varying specific C~{\sc iii} $\lambda$977 flux
(i.e.  flux in active regions computed through an empirical relationship
between the filling factor and $P_{\rm rot}$). The plots stop at the rotation
period of the Sun because of increasing uncertainties resulting from the
filling factor-P$_{\rm rot}$ relationship. \label{figPhys}} 
\end{figure}

Most notably, the power-law slope for the electron pressure and the C~{\sc iii}
977~\AA\ flux are very similar, namely $\simeq -1.65$. Note, however, that
the flux power-law relationship has been computed from the integrated flux and
it thus represents an average over the stellar surface. If, however, the
emitting TR plasma only occupies a fraction ($A$; $A\le1$) of the surface,
known as filling factor, then the specific flux emitted by the active regions
will be correspondingly larger. The pressure power-law does not depend on the
area factor if only one atmospheric component dominates. Thus, assuming that
the same fraction of C~{\sc iii} 977~\AA\ is emitted over the same $d\ln{T_e}$
interval for our sample then we can adopt
\begin{equation}
{F_{A=1}\over{A}} \propto P_e^2 {dr\over{d\ln T_e}}
\end{equation}
where $F_{A=1}$ represents the surface integrated flux. Using the power-law
relationships derived above, one can write
\begin{equation}\label{eqffact}
{{d\ln T_e}\over{dr}}\propto A\;P_{\rm rot}^{-1.77}
\end{equation}
as a function of the rotation period. Thus, with empirically measured filling
factors for the solar proxies or with a reliable relationship between $A$ and
$P_{\rm rot}$ (or Rossby number\footnote{The Rossby number ($R_{\circ}$), which
is proportional to the rotation period of the star and inversely proportional
to the turnover time at the base of the convective zone, is commonly used as a
measure of the activity level of a star. For our targets, which have very
similar masses and spectral types, one can safely assume that the Rossby number
is simply proportional to $P_{\rm rot}$.}) the relationship above can yield an
estimate of the dependence of the electron temperature gradient with age. For
example, using the empirical fit of Montesinos \& Jordan (1993) that relates
the magnetic filling factor with the Rossby number, we can write for our solar
proxies
\begin{equation}
\log A \simeq -0.86\;R_{\circ} \simeq -0.068\;P_{\rm rot}
\end{equation}
and thus, the temperature gradient can be expressed as
\begin{equation}
{{d\ln T_e}\over{dr}}\propto P_{\rm rot}^{-1.77} \; 10^{-0.068\;P_{\rm rot}}
\end{equation}
A representation of the relative variations of the temperature gradient, the
electron pressure and the specific C~{\sc iii} $\lambda$977 flux (i.e., flux
divided by filling factor) are provided in Figure \ref{figPhys}. Note that
these relationships are only represented for rotation periods shorter than that
of the Sun, where the filling factor is better defined. Strikingly, the
specific C~{\sc iii} $\lambda$977 flux remains relatively constant with
rotation period: as the filling factor increases, the total stellar flux
increases correspondingly. However, since the electron density increases with
decreasing rotation period, the temperature gradient must also increase to keep
the specific flux relatively constant. It is worth noting that, given the large
change in the implied temperature gradient across the sample, one or more of
the several assumptions on which Eq. \ref{eqffact} is based may no longer hold.

\section{FUV irradiances and effects on paleo-planetary atmospheres}

The relationships obtained from the solar proxies indicate that the total solar
FUSE/FUV flux was about twice the present value 2.5 Gyr ago and about 4 times
the present value about 3.5 Gyr ago. Note also that the FUSE/FUV flux of the
Zero-Age Main Sequence Sun could have been stronger than today by as much as 50
times.

As discussed by Canuto et al. (1982, 1983), Luhmann \& Bauer (1992), Ayres
(1997), Guinan, Ribas \& Harper (2002) and others, the strong FUV and UV
emissions of the young, more active Sun could have played a major role in the
early development and evolution of planetary atmospheres -- especially those
of the terrestrial planets. The expected strong X-ray--UV irradiance of the
young Sun can strongly influence the photochemistry and photoionization (and
possible erosion) of the early planetary atmospheres and even surfaces (in the
case of Mercury, Moon and Mars) and also may play a role in the origin and
development of life on Earth as well as possibly on Mars. For example, Canuto
et al. (1982, 1983) discuss the photochemistry of O$_2$, O$_3$, CO$_2$, H$_2$O,
etc, in the presumed outgassed CO$_2$-rich early atmosphere of the Earth. Ayres
(1997) discusses the effect of the young Sun's increased ionizing X-ray--UV
flux, and possible accompanying enhanced solar wind, on the erosion of the
early atmosphere of Mars about 3--4 Gyr ago. Also, Lammer et al. (2003) utilize
irradiance data from the ``Sun in Time'' project and solar-wind estimates by
Wood et al. (2002a) to evaluate the mechanisms for loss of water from Mars and
study the implications for the oxidation of the Martian soil.

Similarly, our data can also provide insights into the so-called Faint Sun
Paradox. The paradox arises from the fact that standard stellar evolutionary
models show that the Zero-Age Main Sequence Sun had a luminosity of $\sim$70\%
of the present Sun. This should have led to a much cooler Earth in the past
while geological and fossil evidence indicate otherwise. A solution to the
Faint Sun Paradox proposed by Sagan \& Mullen (1972) was an increase of the
greenhouse effect for the early Earth. The gases that have been suggested to
account for this enhanced greenhouse effect are carbon dioxide (CO$_2$),
ammonia (NH$_3$) or methane (CH$_4$). However, recent results for atmospheric
composition of the early Earth (Rye, Kuo, \& Holland 1995) are in conflict with
the high levels of CO$_2$ and H$_2$O needed to explain the stronger greenhouse
effect.  Moreover, ammonia is a likely candidate except that it is quickly and
irreversibly photodissociated. However, Sagan \& Chyba (1997) proposed that
hydrocarbon-based aerosols could shield the ammonia from damaging incoming
solar radiation. Another alternative explanation for the enhanced greenhouse
effect is from atmospheric methane as discussed by Pavlov et al. (2000).
Methane is a very strong greenhouse gas (10$^3$--10$^4$ times stronger than
CO$_2$) and the necessary amount of CH$_4$ could have been provided by
methanogenic bacteria in the proposed early Earth's reducing atmosphere. 

In addition to the stronger greenhouse effect, the expected higher X-ray--UV
irradiance of the young Sun should heat up the upper atmosphere of the early
Earth. Although the stronger high-energy solar radiation cannot by itself
explain the Faint Sun Paradox, the photoionization and photodissociation
reactions triggered could play a major role in what greenhouse gases are
available. For example, the high levels of FUV-UV radiation of the young Sun
could strongly influence the abundances of ammonia and methane in the
pre-biotic and Archean planetary atmosphere some 2-4 Gyr ago. Similarly, the
photochemistry and abundance of ozone (O$_3$) is of great importance to study
life genesis in the Earth. Ozone is an efficient screening mechanism for the
enhanced UV radiation of the young Sun, thus protecting the emerging life on
the Earth's surface.

To fully evaluate the influence of FUV radiation in the developing planetary
system one must first estimate the flux contribution of the strong H~{\sc i}
Ly$\alpha$ feature, which is not included in the FUSE wavelength range.
Preliminary estimates using HST/STIS spectra indicate that Ly$\alpha$
$\lambda$1216 contributes a significant fraction (up to 90\%) of the total FUV
flux. Further observations and measurements in the near future will allow us to
obtain this missing piece of information and complete the FUV irradiance study.

\section{Conclusions}

In this study we have utilized FUV spectra acquired with the FUSE satellite to
investigate the FUV emission characteristics of solar analogs. Our six targets
are well-known G0--5 solar-type stars especially selected to serve as proxies
for the Sun at different ages, nearly covering the entire solar main sequence
lifetime from 130 Myr to $\sim$9 Gyr. Here we have focused on two aspects of
the ``Sun in Time'' program which are the study of TR plasma electron pressures
(using C~{\sc iii} $\lambda$1176/$\lambda$977 line ratio diagnostics) for stars
that differ only in rotation period (and age) and the evolution of irradiances
for specific features covering emitting plasma with temperatures from
$\sim10^4$~K to $\sim10^7$ K.

To analyze the plasma density of the TR we have used a theoretical relationship
between the ratio of C~{\sc iii} $\lambda$1176 and $\lambda$977 emission line
fluxes and the electron pressure ($P_e$) of the material responsible for the
emission. Fortunately, both C~{\sc iii} transitions are within the FUSE
wavelength range so that they can be measured simultaneously. Our results
indicate a power-law relationship between the electron pressure and the stellar
rotation period and overall magnetic-related activity (both related to age).
The slope of this relationship has been found to be $\sim$$-$1.7, which
suggests that the electron pressure of $\sim$10$^5$-K material in the Sun has
decreased significantly since the beginning of its main sequence evolution. The
higher values of electron pressure found for the more active stars are best
explained by stronger magnetic confinment of the TR plasma.

The measured fluxes for four emission features -- C~{\sc ii} $\lambda$1037,
C~{\sc iii} $\lambda$977, O~{\sc vi} $\lambda$1032, and Fe~{\sc xviii}
$\lambda$975 -- were referred to a radius of 1~$R_{\odot}$ and also corrected
for ISM absorption whenever necessary. The typical formation temperatures of
the studied features are $20,000$~K, $60,000$~K, $300,000$~K, and 6 MK,
respectively. Our analysis indicates that the evolution of these fluxes with
stellar rotation period or age can be accurately fit with power-law
relationships of different slopes. Interestingly, the slopes not only change
from line to line, but also they show a clear trend: The power law becomes
steeper as we move from cooler to hotter plasmas, with the most extreme trend
being that of the coronal ion Fe~{\sc xviii}. Also, the evolution total
integrated flux in the FUSE wavelength range (920--1180~\AA) can be fit with a
power-law relationship with a slope of $\sim$$-$1.7, which indicates a factor
of $\sim$50 flux decrease along the solar main sequence evolution as the
magnetic dynamo activity decreases.

The high levels of FUV line emission fluxes (and related high-energy emission)
of the early Sun could have played a crucial role in the photochemistry and
photoionization of terrestrial and planetary atmospheres. To address this point
we are completing spectral irradiance tables covering 1~\AA\ to 3200~\AA\ for
our program stars that can serve as input data for evolution and structure
models of the paleo-atmospheres of the Solar System planets. The FUSE
observations fill a critical wavelength and energy gap in the ``Sun in Time''
program and complement observations of the same stars in the X-ray and EUV
regions (corona) made with ROSAT, SAX, ASCA, XMM, Chandra, and EUVE, and in the
UV (TR and chromosphere) made using IUE and HST. The only missing piece of
information is the evolution of the irradiance of the strong chromospheric
H~{\sc i} Ly$\alpha$ $\lambda$1216 FUV feature. We are currently carrying out
this part of the study and we expect to complete it shortly. Full spectral
irradiance tables for five solar proxies are thus nearing completion and will
be made available in a forthcoming publication (Ribas \& Guinan 2003, in
preparation).

We anticipate that these irradiance results will be important for the study of
paleo-atmospheres of the Solar System planets. In particular, preliminary
analyses indicate that the high X-ray and EUV emission fluxes of the early Sun
could have produced significant heating of the planetary exospheres and
upper-atmospheres thus enhancing processes such as thermal escape. The early
Sun's strong FUV and UV fluxes penetrate further into the atmosphere and
probably influenced the photochemistry of, e.g., methane and ammonia, which are
important greenhouse gases. Thus, this study has strong implications for the
evolution of the pre-biotic and Archean atmosphere of the Earth as well as for
the early development of life on Earth and possibly on Mars.

\acknowledgments

We thank Seth Redfield for making available to us the data from the Colorado
LISM model, which we have used to correct the measured fluxes for ISM
absorption. The referee, Tom Ake (Johns Hopkins Univ.), is thanked for helpful
comments and suggestions that led to significant improvements. We acknowledge
with gratitude the support for the ``Sun in Time'' program from NASA-FUSE
grants NAG 5-08985, NAG 5-10387, NAG 5-12125 and also from NSF-RUI grant
AST-00-71260. G.M.H.'s research was funded by NASA grant NAG5-4808 (LTSA).
This research has made use of the SIMBAD database, operated at CDS, Strasbourg,
France.


\begin{deluxetable}{lrlrcccccl}
\tablewidth{0pt}
\tablefontsize{\footnotesize}
\tablecaption{Relevant data for the ``Sun in Time'' FUSE targets and the Sun.
\label{tabprop}} 
\tablehead{\colhead{Name} &
\colhead{HD} &
\colhead{Sp.} &
\colhead{d} &
\colhead{$T_{\rm eff}$} &
\colhead{Mass} &
\colhead{Radius} &
\colhead{$P_{\rm rot}$} &
\colhead{Age} &
\colhead{Age}\\
\colhead{} &
\colhead{} &
\colhead{Typ.} &
\colhead{(pc)} &
\colhead{(K)} &
\colhead{(M$_{\odot}$)} &
\colhead{(R$_{\odot}$)} &
\colhead{(d)} &
\colhead{(Gyr)} &
\colhead{indicator}}
\startdata
EK Dra        & 129333 & G0~V  &33.9 &5818&1.07&0.95&2.75    &0.13   & Pleiades str.\\
$\pi^1$~UMa   &  72905 & G1.5~V&14.3 &5840&0.98&0.97&4.68    &0.3    & UMa str.\\
$\kappa^1$ Cet&  20630 & G5~V  & 9.2 &5700&1.01&0.94&9.2     &0.65   & $P_{\rm rot}$-Age rel.\\
$\beta$ Com   & 114710 & G0~V  & 9.2 &5950&1.11&1.08&12.4    &1.6    & $P_{\rm rot}$-Age rel.\\
Sun           &  --    & G2~V  & 1 AU&5777&1.00&1.00&25.4    &4.6    & Isotopic dating\\
$\beta$ Hyi   &   2151 & G2~IV & 7.5 &5800&1.09&1.88&$\sim$28&6.7    & Isochrones\\
16 Cyg A      & 186408 & G2~V  &21.6 &5740&0.99&1.27&$\sim$35&$\sim$9& Isochrones\\
\enddata
\end{deluxetable}

\begin{deluxetable}{lclccrrrrr}
\tablewidth{0pt}
\tablefontsize{\scriptsize}
\tablecaption{Measured Integrated Fluxes for Selected Emission
Features\tablenotemark{a} \label{taball}}
\tablehead{\colhead{Name} & 
\colhead{Obs ID} &
\colhead{Date Obs.} &
\colhead{t$_{\rm exp}$ (ks)} &
\colhead{${\cal F}^{\mbox{\tiny C{\sc iii}}}_{\mbox{\tiny $\lambda$977}}$} & 
\colhead{${\cal F}^{\mbox{\tiny C{\sc iii}}}_{\mbox{\tiny $\lambda$1176}}$} & 
\colhead{${\cal F}^{\mbox{\tiny O{\sc vi}}}_{\mbox{\tiny $\lambda$1032}}$}& 
\colhead{${\cal F}^{\mbox{\tiny O{\sc vi}}}_{\mbox{\tiny $\lambda$1038}}$}& 
\colhead{${\cal F}^{\mbox{\tiny C{\sc ii}}}_{\mbox{\tiny $\lambda$1037}}$}& 
\colhead{${\cal F}^{\mbox{\tiny Fe{\sc xviii}}}_{\mbox{\tiny $\lambda$975}}$}}
\startdata
EK Dra        & C1020501 & 2002 May 14 & 24.2&11.1$\pm$0.7 & 7.6$\pm$0.4 & 7.3$\pm$0.3 & 3.4$\pm$0.2 &0.47$\pm$0.07&   0.81$\pm$0.19\\
$\pi^1$~UMa   & B0780101 & 2001 Dec 5  & 16.5&14.7$\pm$0.7 & 9.3$\pm$0.5 & 9.6$\pm$0.3 & 4.8$\pm$0.2 &0.74$\pm$0.09&   0.65$\pm$0.26\\
$\kappa^1$ Cet& A0830301 & 2000 Sep 10 & 13.0&16.4$\pm$0.7 &10.7$\pm$0.4 &12.4$\pm$0.4 & 6.2$\pm$0.2 &1.07$\pm$0.12&   0.44$\pm$0.14\\
$\beta$ Com   & A0830401 & 2001 Jan 26 & 15.0&11.0$\pm$0.5 & 5.5$\pm$0.2 & 5.9$\pm$0.2 & 2.8$\pm$0.1 &0.63$\pm$0.08&$<$0.06\tablenotemark{b}\\
$\beta$ Hyi   & A0830101 & 2000 Jul 1  & 18.0&15.1$\pm$0.5 & 5.8$\pm$0.2 & 6.1$\pm$0.2 & 2.8$\pm$0.1 &1.02$\pm$0.09&$<$0.04\tablenotemark{b}\\
16 Cyg A      & C1020101 & 2002 Jul 30 & 35.0&0.66$\pm$0.14&0.20$\pm$0.09&0.43$\pm$0.05&0.20$\pm$0.03&$<$0.05\tablenotemark{b}&$<$0.04\tablenotemark{b}\\
\enddata
\tablenotetext{a}{Fluxes in units of $10^{-14}$ erg s$^{-1}$ cm$^{-2}$}
\tablenotetext{b}{1-$\sigma$ upper limits}
\end{deluxetable}

\begin{deluxetable}{lrrrrrrrrr}
\tablewidth{0pt}
\tablefontsize{\footnotesize}
\tablecaption{Radial Velocity Measurements for Selected Features in the
FUSE Spectra\tablenotemark{a}
\label{tabRV}}
\tablehead{\colhead{Name}&
\colhead{Phot}&
\multicolumn{2}{c}{C{\sc ii}$_{\mbox{\tiny $\lambda$1037}}$}&
\multicolumn{2}{c}{C{\sc iii}$_{\mbox{\tiny $\lambda$977}}$}&
\multicolumn{2}{c}{O{\sc vi}$_{\mbox{\tiny $\lambda$1032}}$}&
\multicolumn{2}{c}{Fe{\sc xviii}$_{\mbox{\tiny $\lambda$975}}$}\\
\colhead{}& 
\colhead{Hel} & 
\colhead{Hel} & \colhead{Shift} &
\colhead{Hel} & \colhead{Shift} &
\colhead{Hel} & \colhead{Shift} &
\colhead{Hel} & \colhead{Shift}}
\startdata    
EK Dra        &$-$30.5&$-$23        &    8         &$-$15 &   16 &$-$25 &    6 &$-$26         &    5         \\
$\pi^1$~UMa   &$-$12.0&$-$23        &$-$11         &$-$24 &$-$12 &$-$19 & $-$7 &$-$51\rlap{:} &$-$39\rlap{:} \\
$\kappa^1$ Cet&   19.9&   12        & $-$8         &   16 & $-$4 &   16 & $-$4 &    1         &$-$19         \\
$\beta$ Com   &    6.1&    3        & $-$3         &    0 & $-$6 &    4 & $-$2 &   --         &   --         \\
$\beta$ Hyi   &   22.7&    6        &$-$17         &   20 & $-$3 &   16 & $-$7 &   --         &   --         \\
16 Cyg A      &$-$25.6&$-$40\rlap{:}&$-$14\rlap{:} &$-$29 & $-$3 &$-$22 &    4 &   --         &   --         \\
\enddata
\tablenotetext{a}{All velocities in units of km~s$^{-1}$.
Phot=stellar photosphere; Hel=heliocentric velocity;
Shift=Doppler shift with respect to stellar photosphere.}
\end{deluxetable}

\begin{deluxetable}{lllllr}
\tablewidth{0pt}
\tablefontsize{\footnotesize}
\tablecaption{ISM- and Radius-corrected Surface Fluxes for the Five of the
Targets with High S/N Data\tablenotemark{a} 
\label{tabsurf}}
\tablehead{\colhead{Name} & 
\colhead{$f^{\mbox{\tiny C{\sc ii}}}_{\mbox{\tiny $\lambda$1037}}$}&
\colhead{$f^{\mbox{\tiny C{\sc iii}}}_{\mbox{\tiny $\lambda$977}}$}&
\colhead{$f^{\mbox{\tiny O{\sc vi}}}_{\mbox{\tiny $\lambda$1032}}$}&
\colhead{$f^{\mbox{\tiny Fe{\sc xviii}}}_{\mbox{\tiny $\lambda$975}}$}&
\colhead{$f^{\mbox{\tiny FUV}}_{\mbox{\tiny 920-1180}}$}}
\startdata
EK Dra        &1.17$\pm$0.18  &28.4$\pm$1.7   &18.2$\pm$0.8   &2.02$\pm$0.51  &-- \\ 
$\pi^1$ UMa   &0.32$\pm$0.04  &6.52$\pm$0.33  &4.10$\pm$0.12  &0.28$\pm$0.11  &  45$\pm$20\\  
$\kappa^1$ Cet&0.20$\pm$0.02  &3.30$\pm$0.15  &2.32$\pm$0.07  &0.082$\pm$0.025&16.4$\pm$3.8\\  
$\beta$ Com   &0.090$\pm$0.011&1.64$\pm$0.07  &0.842$\pm$0.030&$<$0.0085      &-- \\  
$\beta$ Hyi   &0.032$\pm$0.003&0.543$\pm$0.019&0.190$\pm$0.007&$<$0.0012      &-- \\  
16 Cyg A      &   $<$0.03     &0.423$\pm$0.090&0.246$\pm$0.028&$<$0.02        &-- \\
\enddata
\tablenotetext{a}{Surface fluxes in units of $10^4$ erg s$^{-1}$ cm$^{-2}$}
\end{deluxetable}

\begin{deluxetable}{lcll}
\tablewidth{0pt}
\tablefontsize{\footnotesize}
\tablecaption{Flux decrease power law slopes for different wavelength regions
and ions.
\label{tabslopes}}
\tablehead{\colhead{Wav. region/ion} & 
\colhead{Plasma $T$ (K)} & 
\colhead{Slope ($P_{\rm rot}$)\tablenotemark{a}} &
\colhead{Slope (age)\tablenotemark{b}}}
\startdata
FUV (920--1180 \AA)                   &$\sim10^4-10^5$&$-$1.74&$-$1.03\\
C~{\sc ii}                            &$2\cdot10^4$   &$-$1.49&$-$0.88\\
C~{\sc iii}                           &$6\cdot10^4$   &$-$1.61&$-$0.95\\
O~{\sc vi}                            &$3\cdot10^5$   &$-$1.82&$-$1.07\\
X-rays (0.1--2.4 keV)\tablenotemark{c}&$2\cdot10^6$   &$-$2.64&$\;\;\; -$  \\
Fe~{\sc xviii}                        &$6\cdot10^6$   &$-$3.2 &$-$1.9 \\
\enddata
\tablenotetext{a}{Power law slope corresponding to the following functional
form: ${\rm flux} = {\rm flux}_{\circ} \; (P_{\rm rot})^{\alpha}$, where $\alpha$
is the exponent listed in the column}
\tablenotetext{b}{Power law slope corresponding to the following functional
form: ${\rm flux} = {\rm flux}_{\circ} \; ({\rm age})^{\beta}$, where $\beta$
is the exponent listed in the column}
\tablenotetext{c}{From G\"udel et al. 1997}
\end{deluxetable}

\begin{deluxetable}{lccl}
\tablewidth{0pt}
\tablefontsize{\footnotesize}
\tablecaption{C {\sc iii} $\lambda$1176/$\lambda$977 flux ratio and
electron pressure of the transition region
\label{tabR}}
\tablehead{\colhead{Name} & 
\colhead{${\cal R}$\tablenotemark{a}} &
\colhead{${\cal R\tablenotemark{b}}_{\mbox{\tiny ISM}}$} &
\colhead{$\log P_e\;(\mbox{cm}^{-3}\,\mbox{K})$}}
\startdata
EK Dra        &0.68$\pm$0.05&0.67$\pm$0.05&$\ga$16.0\\
$\pi^1$~UMa   &0.63$\pm$0.05&0.61$\pm$0.04&$15.4^{+0.6}_{-0.2}$\\
$\kappa^1$ Cet&0.65$\pm$0.04&0.60$\pm$0.04&$15.3^{+0.5}_{-0.2}$\\
$\beta$ Com   &0.50$\pm$0.03&0.48$\pm$0.03&$14.8^{+0.1}_{-0.1}$\\
$\beta$ Hyi   &0.38$\pm$0.02&0.33$\pm$0.02&$14.3^{+0.05}_{-0.05}$\\
16 Cyg A      &0.30$\pm$0.15&0.27$\pm$0.13&$14.1^{+0.4}_{-0.4}$\\
\enddata
\tablenotetext{a}{${\cal R}\equiv{\cal F}^{\mbox{\tiny C{\sc iii}}}_{\mbox{\tiny
$\lambda$1176}}/ {\cal F}^{\mbox{\tiny C{\sc iii}}}_{\mbox{\tiny
$\lambda$977}}$}
\tablenotetext{b}{C {\sc iii} $\lambda$977 flux corrected for ISM absorption
(see text)} 
\end{deluxetable}


\end{document}